\documentstyle[12pt]{article}
\begin{document}

\title {Superspace without torsion and the composite fundamental fermions}
\author {Michael A. Ivanov \\
Physics Dept.,\\
Belarus State University of Informatics and Radioelectronics, \\
6 P. Brovka Street,  BY 220027, Minsk, Republic of Belarus.\\
E-mail: ivanovma@gw.bsuir.unibel.by}
\date{} \maketitle
\begin{abstract}

(N=2)-superspace without torsion is described,  which is
equivalent to an 8-space with a discrete internal subspace. A
number and a character of ties determine now an internal symmetry
group, while in the supersymmetrical models this one is determined
by an extension degree N. Such a model can be constructed for no
less than 4 generations of the two-component fundamental fermions.
Analogues of the Higgs fields appear in the model naturally after
transition to the Grassmannian extra coordinates. The connection
between discrete and continues internal symmetries of the model is
discussed. If one considers gravity as embedding the curved
4-space into a 12-dimensional flat space, a $U(1)$-symmetry
appears transformations of which should be connected with the ones
of $SU(2)$-group. If super-strong interacting gravitons are
constituents of the composite fermions, all this may open us
another way to unify the known interactions. The main feature of
this new approach may be the external see of gravitons underlying
an internal structure of particles; the lack of any divergencies
would be due the Planckian spectrum of external gravitons.
\end{abstract}

\section[1]{Introduction }
We speak in different languages to describe the unique nature. To
avoid divergencies in the theory one introduces supersymmetry. But
any super-partner of existing particles is unknown. To rich the
accordance with phenomenology, we introduce postulates about the
internal symmetry group - but it is reasonable if the fundamental
particles have some structure. I would like to tell here about a
possibility to tie a few different approaches in particle physics
by means of introduction of a superspace without torsion (see
\cite{201}). Of course, there is not any symmetry between fermions
and bosons in this case. As it was shown by the author, the
discrete internal subspace of the two-component fundamental
fermions may be replaced with a super-subspace of the Grassmannian
extra coordinates in which a set of supermanifolds has been picked
out. An extension with $N=2$ as minimum is necessary to provide
non-zero values of relative coordinates of the composite fermions.
The main goal of the supersymmetrical models - the lack of
divergencies - perhaps, may be achieved in another way: if we
consider super-strong interacting gravitons as constituents of the
composite fermions.

\section[2]{An internal space of the two-component fermions}
Internal coordinates $y_{A}^{\mu}$ of the two-component fermions
in the model \cite{201} should obey the condition:
\begin{equation}
y_{A}^{\mu}=0 \bigvee y_{A}^{\mu}=\pm l,\ A=2,3,4;\ 6,7,8;
\end{equation}
where $l$ is a constant of the length dimension. It means that the
quark sector of the model is characterized by one length $l.$ This
condition does not affect the values of $y_{A}^{\mu}$ for $A=1,5,$
and one has five constants with such dimension (so as
$y_{1}^{\mu}=-y_{5}^{\mu})$ in the model, four of which concern
the lepton sector only. \par It was shown by the author that the
discrete symmetries of the model, which result from the field
equations structure in an 8-space, may be interpreted by an
observer in the Minkowski space as the continues internal
symmetries \cite{201, 202}. They are connected with mixing of
different solutions of field equations distinguished with internal
coordinate values only. This algebraic structure is such the one
that $SU(3)_{c}\times SU(2)_{l}$ is the global symmetry group of
the model with four genarations. A permissible multiplet of the
group differs from the one taken in the standard model only by an
existence of another $SU(2)_{r}-$singlet for states with $A=1,5$
(in the lepton sector) that may be used to provide a non-zero mass
of neutrino. After the observation of evidence of non-zero
neutrino mass difference by the Super-Kamiokande collaboration
\cite{100}, my model is more actual than in 1990. It was shown in
\cite{201} that transition from the internal coordinates $y^{\mu}$
to the Grassmannian ones $\chi_{a},\bar{\chi}_{a},\ a=1,2,3,4,$
leads to the appearance of the effective four-dimensional fields
in the model's Lagrangian. The ones are $SU(2)-$doublets and
analogs of the Higgs fields and may play their part in the
mass-splitting mechanism. \par The vectorial $y_{\mu}$ and spinor
$\chi_{a},\bar{\chi}_{a}$ coordinates are connected between
themselves:
\begin{equation}
y^{\mu}=\bar{\chi}\gamma^{\mu}\chi,
\end{equation}
$\gamma^{\mu}$ are the Dirac matrices. One must consider
$\chi_{a},\bar{\chi}_{a},$ to be independent of the Minkowskian
coordinates $x^{\mu},$ then $\chi_{a},\bar{\chi}_{a}$ may commute
or anticommute between themselves. In last case,
$(x,\bar{\chi},\chi)$ is the $(N=2)-$superspace without torsion.
If $\chi_{a},\bar{\chi}_{a}$ anticommute, the additional condition
must take place:
\begin{equation}
R^{2}+8M =0,
\end{equation}
where $R=\bar{\chi}\gamma^{5}\chi,\
\gamma^{5}=i\gamma^{0}\gamma^{1}\gamma^{2}\gamma^{3},$
$M=\bar{\chi}_{1}\chi_{1}\bar{\chi}_{4}\chi_{4}+(\bar{\chi}_{2}\chi_{2}+
\bar{\chi}_{1}\chi_{1})\bar{\chi}_{3}\chi_{3}
+\bar{\chi}_{1}\chi_{4}\bar{\chi}_{2}\chi_{3}+
\bar{\chi}_{3}\chi_{2}\bar{\chi}_{4}\chi_{1}.$ The set of ties (2)
picks out $8$ supermanifolds; you can name them "branes" if you
want. They are "compactified" in the very simple and definite
manner. But in another language, we deal here with the discrete
internal subspace. \par In the model's Lagrangian, which must
contain terms of expansion on degrees of $\bar{\chi},\chi,$
factors by the first degree of $\chi_{A}$ should be sets of
scalars with respect to the Lorentz group and $SU(2)-$doublets,
i.e. they will be analogs of the Higgs fields of the standard
model.

\section[3]{Gravitons as possible constituents of the composite fermions}
It was also shown by the author \cite{203} that embedding the
general relativity $4-$space into a flat $12-$space gives an
alternative model of gravitation with a global $U(1)-$symmetry and
the discrete $D_{1}-$one. The last one may lead to the
$SU(2)-$symmetry of the unified model. It is an exciting puzzle:
may these two $SU(2)-$symmetries - from \cite{201} and \cite{203}
- be identified or not? \par The Ricci tensor $r_{\mu \nu}$ in the
model \cite{203} is equal to:
\begin{equation}
r_{\mu \nu} = 2(f^{2} - f) \tilde \Gamma^{\alpha}_{\epsilon [
\alpha} \tilde \Gamma^{\epsilon}_{\nu ]  \mu},
\end{equation}
where $f$ is a free parameter, and $\tilde \Gamma$ is the
connection.  The parameter $f$ can have any value, excluding $f=0;
1.$ On the manifold $\Sigma^{4},$ the global variations of $f$ are
not observable, and $U(1)$ will be the global symmetry group of
the model. The discrete $D_{1}- $symmetry will take place on
$\Sigma^{4}$ due to the quadratic dependence of $r_{\mu \nu}$ on
$f$. The last symmetry may be transformed into the global
$SU(2)$-symmetry. But a variation of the parameter $f$ can lead to
a permutation of a pair of solutions which will be transformed by
the group $SU(2).$ To avoid this dependence of two
transformations, one can perform a rotation in the plane $(f,F)$
(where $F=f^{2}-f$) on some angle $\theta.$ Under an additional
condition that one component of the $SU(2)$-doublet should be
almost massless after breaking of the $SU(2)$-symmetry (i.e. $f
\to 1$ for this component), $\theta$ has the minimum value $\theta
_{min}$ with $\sin^{2}\theta _{min} =0.20$; here $\theta_{min}$ is
an analog of the Weinberg angle of the standard model.

\par
A quantum mechanism of classical gravity based on an existence of
the external sea of super-strong interacting gravitons was
described by the author for the Newtonian limit \cite{204}. This
mechanism needs graviton pairing and "an atomic structure" of
matter for working it. In this approach, the two fundamental
constants - Hubble's and Newton's ones - are connected between
themselves (for more details about confrontation of this model
with observations, see \cite{205}). If this low-energy quantum
gravity is adequate to the nature (an effective temperature of the
graviton background must be the same as of CMB) then gravitons
might be constituents of the composite fermions. The dimensional
constant $D$ of this model is equal to: $D=1.124 \cdot
10^{-27}{m^{2} / eV^{2}},$ so that for forehead collisions of any
two particles with energies $E$ and $\epsilon$ we have for the
cross-section $\sigma(E,\epsilon)$: $\sigma(E,\epsilon)= D\cdot
E\cdot\epsilon$. For example, by $E \sim \epsilon \sim 5~ keV,$
such the interaction would have the same intensity as the strong
interaction. But any divergence, perhaps, would be not possible
because of natural smooth cut-offs of the {\it external} graviton
spectrum from both sides.

\section[4]{Conclusion}
It is obvious that there are much more open problems here than
solved ones but some findings may serve as the milestones for
future developments. The considered model of the two-component
fundamental fermions describes four generations, and an appearance
of any particle of the fourth one would be crucial for its
progress. The considered idea of underlying discrete symmetries
leads us deeper to the machinery of "elementary" particles. The
concept of very-low-energy quantum gravity has some support in the
discovery of quantum states of ultra-cold neutrons in the Earth's
gravitational field by Nesvizhevsky's team \cite{206}: observed
energies of levels (it means that their differences, too) have the
order of $10^{-12}$ eV. If the considered quantum mechanism of
gravity is realized in the nature, both the general relativity and
quantum mechanics should be modified.

\end{document}